\documentclass[11pt]{article}
\usepackage{moriond,epsfig}

\def\be{\begin{equation}}
\def\ee{\end{equation}}
\def\bea{\begin{eqnarray}}
\def\eea{\end{eqnarray}}

\begin{document}
\vspace*{4cm}
\title{RELATIVISTIC TREATMENT OF NEUTRINO OSCILLATIONS \\ IN MOVING MATTER}

\author{ A.I. STUDENIKIN}

\address{Department of Theoretical Physics, Moscow State University, \\
119299 Moscow, Russia, e-mail: studenik@srd.sinp.msu.ru}

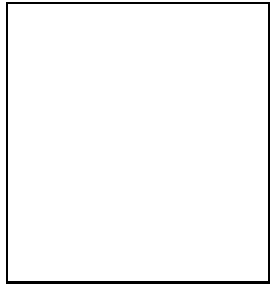
\maketitle\abstracts{A review on the Lorentz 
invariant treatment of neutrino spin 
and flavour oscillations in moving and polarized matter is presented. 
Within this approach it becomes possible to consider
neutrino oscillations in arbitrary electromagnetic 
fields. It is also shown that neutrino effective potential in matter can be 
significantly changed by relativistic motion of matter.}

In this paper I should like to give a short review on the recent 
studies 
of neutrino oscillations in moving and polarized matter, with an 
electromagnetic field being also imposed, that have been performed 
during 
the last few years by our research group at the Moscow State 
University. 
For about six years ago we, while studying neutrino oscillations 
in strong 
magnetic fields, to our much surprise realized that in literature 
there 
were no attempts to consider neutrino spin evolution in any 
electromagnetic 
field rather than constant in time and transversal in respect to 
the 
neutrino propagation magnetic field $\vec B_{\perp}$. The 
influence of the 
longitudinal component of magnetic field was previously neglected 
because 
this component in the rest frame of relativistic neutrino, 
$\vec B^{0}_{\parallel}$, can be suppressed in respect to the 
transversal 
component  which acquire a factor $\gamma=(1-\beta^2)^{-1/2}$ in 
the rest 
frame of the particle: $\vec B^{0}_{\perp}=\gamma \vec B_{\perp}$
($\beta$ is the neutrino speed). 

Further more, in all of the studies of neutrino flavour 
oscillations in 
matter performed before 1995, the matter effect was treated only in 
the nonrelativistic limit that was found to be adequate in 
applications 
to the stellar collapse (see, for example, ref. \cite {FMW87}) 
or to the solar environments accounting for weak currents, ref. \cite 
{HaxZha91}. 

The first our attempt to consider neutrino flavour 
oscillations in 
matter in the case when matter is moving with relativistic 
speed was 
made in 1995, ref. \cite {LikStu94}. In that our study we tried to apply 
the Lorentz invariant formalism for description of neutrino 
flavour oscillations and realized that the value of the matter 
term in the neutrino effective potential can be significantly 
changed if matter is moving with relativistic speed. However, we 
have continued our studies on evaluation of the Lorentz invariant 
formalism in neutrino oscillations only in 1999 ( see ref. \cite 
{ELStpl00} and references therein) with investigation of neutrino 
oscillations in arbitrary electromagnetic fields. We start from 
the Bargmann-Michel-Telegdi (BMT) equation for evolution of the 
spin $S_{\mu}$ of a neutral particle with nonvanishing magnetic, 
$\mu$, and electric, $\epsilon$, dipole moments in electromagnetic 
field, given by its tensor $F_{\mu \nu}=(\vec E, \vec B)$:
\begin{equation}
{dS^{\mu} \over d\tau} =2\mu \Big\{ F^{\mu\nu}S_{\nu} -u^{\mu}(
u_{\nu}F^{\nu\lambda}S_{\lambda} ) \Big\} +2\epsilon \Big\{ 
{\tilde
F}^{\mu\nu}S_{\nu} -u^{\mu}(u_{\nu}{\tilde
F}^{\nu\lambda}S_{\lambda}) \Big\},
\label{1}
\end{equation}
This form of the BMT equation corresponds to the case of the 
particle moving
with constant speed,
$\vec \beta=const$, $u_\mu=(\gamma,\gamma \vec \beta)$.
The spin vector satisfies the usual conditions,
$S^2=-1$ and $S^{\mu}u_\mu =0$.
We generalize the BMT equation for the case of massive 
neutrino by including the effects of weak interactions with 
background matter and  finally arrive 
to the following equation for the 
three dimensional neutrino spin vector $\vec S$:  
\begin{equation} 
{d\vec S \over dt}=  \Big[ {\vec
S} \times \Big({2\mu \over \gamma} {\vec B_0}+{2\epsilon \over 
\gamma}{\vec E_0}+\big(V-{\delta m^{2}A(\theta) \over {2E}}\big){\vec n}\Big) 
\Big], 
\label{BMT}
\end{equation} 
where $\delta m^{2}$ is the mass squared difference and $E$ is the 
energy of neutrino.
The derivative in the left-hand
side of Eq.(\ref{BMT}) is taken with respect to time $t$ in the
laboratory frame, whereas the values $\vec B_0$ and $\vec E_0$ are
the magnetic and electric fields in the neutrino rest frame
\begin{equation} 
\vec B_0=\gamma\Big(\vec 
B_{\perp}
+{1 \over \gamma} \vec B_{\parallel} + \sqrt{1-\gamma^{-2}}
\Big[{\vec E_{\perp} \times \vec n}\Big]\Big), \  
\vec
E_0=\gamma\Big(\vec E_{\perp} +{1 \over \gamma} \vec E_{\parallel} 
-\sqrt{1-\gamma^{-2}}\Big[{\vec B_{\perp} \times \vec
n}\Big]\Big), \ \vec n= {\vec \beta \over \beta},
\label{F0} 
\end{equation} 
where $\vec F_{\perp,\ \parallel}$  ($F=\ B$ or $E$) are fields components in 
the laboratory frame.
The two parameters,
$A(\theta)$ being a function of vacuum mixing angle and
$V=V(n_{eff})$ being the difference of neutrino effective 
potentials
in matter, depend on the nature of neutrino conversion processes in
question.  For specification of $A(\theta)$ and $V(n_{eff})$ 
for different types of
the neutrino conversions see, for example, in 
refs.\cite{LikStu95,AkLaSc97}.  

Using the neutrino spin evolution equation we get the effective 
Hamiltonian that determines the evolution
of the
system $\nu=~(\nu_R,\nu_L)$ in presence of electromagnetic field 
with
given ( in the laboratory frame) components $B_{\parallel, 
\perp}(t)$,
$E_{\parallel, \perp}(t)$:
\begin{equation}
H=({\vec \sigma}{\vec n}) \Bigg({\delta m^2A(\theta) \over
4E}-{V \over 2}-{{\mu B_{\parallel} + \epsilon
E_{\parallel}} \over \gamma}\Bigg) -\mu{\vec \sigma}\Big({\vec B}_{\perp}
+\big[{{\vec E_{\perp}} \times {\vec n}}\big]\Big)-
\epsilon{\vec
\sigma}\Big({\vec E}_{\perp} -\big[{{\vec B_{\perp}} \times {\vec
n}}\big]\Big)+ O({1 \over \gamma^{2}}).
\label{H} \end{equation}
This Hamiltonian accounts for the direct interaction of 
neutrino with electromagnetic fields, 
for transversal fields it reproduces the result 
of ref. \cite {VolVysOku86}. There could be also indirect 
influence of electromagnetic field on neutrino due to the matter 
polarization  by 
the longitudinal magnetic field of the considered electromagnetic 
field configuration ( see ref. \cite {NuSeSmVa97} and references therein). 
The former effect is included into the 
difference of neutrino potentials in matter $V$. 

The obtained Hamiltonian enables us to consider neutrino spin 
procession in an arbitrary configuration of electromagnetic 
fields, including those that contains strong
longitudinal components, and derive the corresponding resonance 
conditions for neutrino oscillations. As one of the examples, let 
us consider the new effect of the
neutrino spin conversion $\nu_{L} \leftrightarrow \nu_{R}$ that 
could appear when neutrinos propagate
in matter under the influence of a field of electromagnetic wave 
and a constant longitudinal magnetic field superimposed. We 
suppose that the neutrino speed is constant and
denote by the unite vector ${\vec e}_{3}$ the axis that is 
parallel with ${\vec n}$
and by $\phi$ the angle between ${\vec e}_{3}$ and the direction 
of
the wave propagation (for simplicity we shall neglect terms
proportional to the neutrino electric dipole moment $\epsilon$). 
In
this case the magnetic field in the neutrino rest frame is given 
by
\begin{equation} 
{\vec B_{0}}=~\gamma\Big[B_1
(\cos{\phi}-\beta){\vec e}_{1} +B_2 (1-\beta\cos{\phi}){\vec e}_2 
-{1
\over \gamma}B_1\sin{\phi}{\vec e}_{3}\Big], \label{B}
   \end{equation} 
where ${\vec e}_{1,2,3}$ are the unit orthogonal
vectors. 
For the electromagnetic wave of circular polarization
propagating in matter it is easy to get:  
\begin{equation}
B_1=B\cos{\psi}, ~B_2=B\sin{\psi},
\label{B12} 
\end{equation} 
where $B$ is the amplitude 
of
the magnetic field in the laboratory frame and the phase of the 
wave
at the point where the neutrino is located at given time $t$ is
\begin{equation} \psi=g\omega t\Big(1-{\beta \over
\beta_{0}}\cos{\phi}\Big),  \label{B12_} \end{equation} 
where $\omega$ is the electromagnetic wave frequency.
The phase depends on the wave speed $\beta_{0}$ in matter 
($\beta_{0}\leq 1$). The values $g=\pm 1$ correspond to the two 
types of the circular polarization of the wave.

In the adiabatic approximation the probability
of neutrino conversion $\nu_L \rightarrow \nu_R$ can be written in 
the form ($x$ is the distance travelled by neutrino),
\begin{equation}
P_{\nu_L \rightarrow \nu_R} (x)=\sin^{2} 2\theta_{eff} 
\sin^{2}{\pi x 
\over L_{eff}},\ \ \ 
sin^{2} 2\theta_{eff}={E^2_{eff} \over 
{E^{2}_{eff}+\Delta^{2}_{eff}}}, \ \ \ 
L_{eff}={2\pi \over \sqrt{E^{2}_{eff}+\Delta^{2}_{eff}}},
\end{equation}
where 
\begin{equation}
E_{eff}=2\mu B (1 -\beta\cos{\phi})
\end{equation}
(terms $\sim O(\gamma^{-1})$ are ommited  here),
and 
\begin{equation}
\Delta_{eff}= V  -{\delta m^2A(\theta) \over 2E} -g\omega\Big(1-{\beta \over 
\beta_{0}}\cos\phi\Big) + 2{{\mu B_\parallel} \over {\gamma}}.
\end{equation}
The corresponding resonance condition now is:
\begin{equation}
V  -{\delta m^2A(\theta) \over 2E} -g\omega\Big(1-{\beta \over 
\beta_{0}}\cos\phi\Big) +2{{\mu B_\parallel} \over {\gamma}} =0.  
\label{Res_3} \end{equation}
Thus, we predict the new
type of resonances in neutrino oscillations $\nu_L 
\leftrightarrow \nu_R$
that can exist in presence of the combination of electromagnetic 
wave and constant longitudinal magnetic field. It is easy to see 
that the longitudinal constant magnetic field can be switched off 
and similar analysis of neutrino oscillations in the field of 
circularly polarized electromagnetic wave is straightforward. 
Neutrino oscillations in the field of linearly polarized electromagnetic wave, 
the possibility of the parametric amplification of neutrino 
oscillations in electromagnetic fields and neutrino spin evolution in 
presence of general external fields were also considered  
in ref. \cite {DvoStu_PAN01}.

Within the developed above approach to neutrino spin evolution
we have focused mainly on description of influence of different
electromagnetic fields, while modeling the matter we confined
ourselves to the most simple case of nonmoving and unpolarized
matter.  Now we should like to go further, ref. \cite {LobStuplb01},
and to generalize the developed Lorentz invariant
approach to the neutrino spin oscillations in arbitrary 
electromagnetic fields for the case of moving and polarized 
homogeneous background matter.
The described below formalism is valid for accounting of matter 
motion and polarization for 
arbitrary ( also relativistic) speed of matter. It should be noted 
here that effects of matter polarization in neutrino oscillations 
were considered previously in several papers 
(see, for example, refs. \cite {NuSeSmVa97,BGN99} and references 
therein). 
However, the used in refs. \cite {NuSeSmVa97,BGN99} procedure of
accounting for the matter polarization effect does not enable one 
to study the case of matter motion with relativistic speed. 
Within our approach we can reproduce corresponding results of refs.  
\cite {NuSeSmVa97,BGN99} in the case of matter which is slowly 
moving or is at rest.

We start again with the BMT spin evolution Eq.(\ref{1}) of
electrodynamics and generalize it for the case when effects of
various neutrino interactions (for example, weak
interaction for which $P$ invariance is broken)
with moving and polarized matter are also taken into account.
The Lorentz invariant
generalization of Eq.(\ref{1}) for our case can be obtained by the 
substitution
of the electromagnetic field tensor
$F_{\mu\nu}=(\vec E,\vec B)$ in the following way:
\begin{equation}
F_{\mu\nu}\rightarrow F_{\mu\nu}+G_{\mu\nu}.
\label{2}
\end{equation}
In evaluation of the tensor $G_{\mu,\nu}$ we demand that the 
neutrino evolution equation must be linear  over the neutrino 
spin, electromagnetic field, the matter fermions currents $j_{f}^{\mu}$ and
polarizations $\lambda_{f}^{\mu}$ ( matter 
is composed of different fermions, $f=\ e,\ n,\ p,\ ...$)
\begin{equation}
j_{f}^\mu=(n_f,n_f\vec v_f), \ \ \ \
\lambda^{\mu}_f =\Bigg(n_f (\vec \zeta_f \vec v_f ),
n_f \vec \zeta _f \sqrt{1-v_f^2}+
{{n_f \vec v_f (\vec \zeta_f \vec v_f )} \over {1+\sqrt{1-
v_f^2}}}\Bigg).
\label{4}
\end{equation}
Here $n_f$, $\vec v_f$, and $\vec \zeta_f \
(0\leq |\vec \zeta_f |^2 \leq 1)$ denote, respectively,
the number densities of the background fermions $f$, the
speeds of the reference frames in which the mean
momenta  of fermions $f$ are zero, and the mean values of 
the polarization
vectors of the background fermions $f$ in the above mentioned 
reference frames. The mean value of the background fermion $f$ 
polarization vector, 
$\vec \zeta_f$, 
is determined by the two-step averaging  
of the fermion relativistic spin 
operator over fermion quantum states in a 
given electromagnetic field and then over fermion statistical 
distribution density function. 
Thus,
in general case of neutrino interaction with different background
fermions $f$ we introduce for description of matter effects
antisymmetric tensor
\begin{equation}
G^{\mu \nu}= \epsilon ^{\mu \nu \rho \lambda}
g^{(1)}_{\rho}u_{\lambda}- (g^{(2)\mu}u^\nu-u^\mu g^{(2)\nu}),
\label{7}
\end{equation}
where
\begin{equation}
g^{(1)\mu}=\sum_{f}^{} \rho ^{(1)}_f j_{f}^\mu
+\rho ^{(2)}_f \lambda _{f}^{\mu}, \ \
g^{(2)\mu}=\sum_{f}^{} \xi ^{(1)}_f j_{f}^\mu
+\xi ^{(2)}_f \lambda _{f}^{\mu},
\label{8}
\end{equation}
(summation is performed over the fermions $f$ of the background). 
The explicit
expressions for the coefficients $\rho_{f}^{(1),(2)}$ and 
$\xi_{f}^{(1),(2)}$
could be found if the particular
model of neutrino interaction is chosen.
In the usual notations the antisymmetric tensor $G_{\mu \nu}$ can 
be
written in the form,
\begin{equation}
G_{\mu \nu}= \big(-\vec P,\ \vec M),
\label{9}
\end{equation}
where
\begin{equation}
\vec M= \gamma \big\{(g^{(1)}_0 \vec \beta-\vec g^{(1)})
- [\vec \beta \times \vec g^{(2)}]\big\}, \
\vec P=- \gamma \big\{(g^{(2)}_0 \vec \beta-\vec g^{(2)})
+ [\vec \beta \times \vec g^{(1)}]\big\}.
\label{10}
\end{equation}
It worth to note that the substitution (\ref{2}) implies that the 
magnetic $\vec B$
and electric $\vec E$ fields are shifted by the vectors $\vec M$ 
and $\vec P$:
$\vec B \rightarrow \vec B +\vec M, \ \ \vec E \rightarrow \vec E -
\vec P$.

We finally
arrive to the following equation for the
evolution of the three-di\-men\-sio\-nal neutrino spin vector 
$\vec S
$ accounting for the direct neutrino interaction with 
electromagnetic
field $F_{\mu \nu}$ and matter (which is described by the tensor
$G_{\mu \nu}$):
\begin{equation}
{d\vec S \over dt}={2\mu \over \gamma} \Big[
{\vec S \times ({\vec B_0}+\vec M_0) \Big]+{2\epsilon \over 
\gamma}
\Big[{\vec S} \times (\vec E_0-\vec P_0)} \Big].
\label{13}
\end{equation}
The influence of matter on the neutrino spin evolution in Eq.(\ref{13}) 
is given by
the vectors $\vec M_0$ and $\vec P_0$ which in the rest frame of 
neutrino
can be expressed in terms of quantities determined in the 
laboratory
frame
\begin{equation}
\vec M_0=\gamma \vec \beta
\Big(g^{(1)}_0-{{\vec \beta \vec g^{(1)}} \over {1+\gamma ^{-
1}}}\Big)
-\vec g^{(1)},\ \ \
\vec P_0=-\gamma \vec \beta
\Big(g^{(2)}_0-{{\vec \beta \vec g^{(2)}} \over {1+\gamma ^{-
1}}}\Big)
+\vec g^{(2)}.
\label{15}
\end{equation}

Let us describe, for example, the electron neutrino propagation in moving and 
polarized electron gas. If we consider the case of the standard 
model supplied with $SU(2)$-singlet right-handed neutrino 
$\nu_{R}$ then neutrino effective interaction Lagrangian reads
\begin{equation}
L_{eff}=-f^\mu \Big(\bar \nu \gamma_\mu {1+\gamma^5 \over 2} \nu 
\Big),\ \ \ 
f^\mu={G_F \over \sqrt2}\Big((1+4\sin^2 \theta _W) j^\mu_e -
\lambda ^\mu _e\Big).
\label{18}
\end{equation}
For the coefficients $\rho_{e}^{(1),(2)}$  we get 
( it is supposed that $\epsilon=0$, so that $\xi^{(i)}_e =0$) 
\begin{equation}
\rho^{(1)}_e={G_F \over
{2\mu \sqrt{2}}}(1+4\sin^2 \theta _W), \ \rho^{(2)}_e
=-{G_F \over {2\mu\sqrt{2}}}.
\label{19}
\end{equation}
Using expressions for the vector $\vec M_0$, Eqs.
(\ref{8}), (\ref{15}), we find,
\begin{equation}
{\vec {M}_0}=\vec\beta\gamma {n_{0} \over \sqrt {1-
v_{e}^{2}}}\Bigg\{
\Big(
\rho^{(1)}+\rho^{(2)}{\vec\zeta_{e}}{\vec v}_e
\Big)
(1-{\vec\beta}{\vec v}_e)
+\rho^{(2)}\sqrt{1-v^2_e}
\Bigg[
{
(\vec \zeta_{e}{\vec v}_e)(\vec\beta{\vec v}_e)
\over
1+\sqrt{1-v^2_e}
}-\vec\zeta_{e}\vec\beta
\Bigg]
+O(\gamma^{-1})
\Bigg\}.
\label{M_0} \end{equation}
It follows that the value of the 
matter effect in neutrino spin evolution depends on the values and 
correlations of the three vectors $\vec \beta, \vec v_{e}$, and $\vec 
\zeta_{e}$. In particular, the matter effect can be "eaten" by the 
relativistic motion of matter if matter is moving along the 
neutrino propagation and $1-\vec \beta \vec v_{e} \approx 0$. 

Now let us discuss, within the Lorentz invariant approach, the neutrino flavour oscillations in moving and 
polarized matter, ref. \cite{GriLobStuplb_02}. 
For simplicity, we consider
neutrino two-flavour oscillations, e.g. $\nu_{e} \leftrightarrow 
\nu_{\mu}$, in matter composed of
only one component, electrons ($f=e$), moving with relativistic 
total speed. Generalizations for
the cases of other types of neutrino conversions and different 
matter compositions and motions are straightforward.

The matter effect in neutrino oscillations occurs as a result of
elastic forward scattering of neutrinos on the background
fermions. In our case the difference $\Delta V$ between the
potentials $V_{e}$ and $V_{\mu}$ for  the two-flavour neutrinos is
produced by the charged current interaction of  the electron
neutrino with the background electrons ( the neutral
current interaction is affective in oscillations between the
active and sterile neutrinos). The corresponding part of the 
neutrino effective
Lagrangian can be written now in the following form 
\begin{equation}
{\cal L}_{eff} =-f^{\mu}\left( {\bar {\nu}}{\gamma}_{\mu}{ 
1+{\gamma}^{5}\over 2}{\nu} \right),\ \ \
f^{\mu}=\sqrt2G_F(j^{\mu}_e-{\lambda}^{\mu}_e).
\end{equation}
This additional term in the Lagrangian modifies the Dirac equation
for neutrino:
\begin{equation} ({\gamma}_0E-{\vec {\gamma}\vec
p}-m)\psi=(\gamma_{\mu} f^{\mu})\psi .
\label{Dir}
\end{equation}
In the limit of weak
potential $|\vec f| \ll p_0=\sqrt{{\vec p}^{\;2}+m^2}$ we
get for the effective energy of the electron neutrino in the
moving and polarized matter
\begin{equation}
E=\sqrt{{\vec p}^{\;2}+m^2}+U \Bigg\{
(1-\vec {\zeta}_e {\vec v}_e)(1-\vec {\beta}{\vec v}_e)+ 
\sqrt{1-v^{2}_e} \Bigg[ \vec
{\zeta}_e {\vec \beta} -{(\vec \beta {\vec v_e})(\vec \zeta_e
{\vec v_e}) \over 1+\sqrt{1-v^2_e}} \Bigg]  \Bigg\} +
O({\gamma}^{-1}),
 \label{energy}
\end{equation}
where in the considered case of the two-flavour neutrino 
oscillations
$\nu_e \leftrightarrow \nu_\mu$ and one-component matter
$U=\sqrt2 G_Fn_0/ \sqrt{1-v^2_e}$, $n_0$ is the invariant matter 
(electron) density.

Thus, in the adiabatic limit the probability
of neutrino conversion $\nu_e \rightarrow \nu_\mu$ can be written 
in the form
\begin{equation}
P_{\nu_e \rightarrow \nu_\mu}(x)=\sin^{2} 2\theta_{eff} \sin^{2} 
{\pi x \over L_{eff}},
\label{ver}
\end{equation}
where the effective mixing angle, $\theta_{eff}$, and 
oscillation length, $ L_{eff}$,
are given by
\begin{equation}
\sin^{2} 2\theta_{eff}={\Delta^{2}\sin^{2} 2\theta \over
{\Big(\Delta \cos2\theta - A\Big)^2+ \Delta^{2}\sin^{2} 2\theta}},
\ \ \ 
L_{eff}= {2\pi \over
{\sqrt {\Big(\Delta \cos2\theta - A\Big)^2+ \Delta^{2}\sin^{2} 
2\theta}}}.
\label{l}
\end{equation}
Here $\Delta = {\delta m^{2}_\nu / {2 |\vec p\,|}}$, $\vec p$ 
is the neutrino momentum, $\theta$ is
the vacuum mixing angle and
\begin{equation}
A=\sqrt2 G_F{n_0 \over \sqrt{1-v^2_e}}\Bigg\{(1-{\vec
\beta}{\vec v_e}) (1-\vec {\zeta}_e {\vec v_e}) + 
\sqrt{1-v^2_e} \Bigg[ \vec {\zeta}_e {\vec
\beta} -{(\vec {\beta} {\vec v_e})(\vec {\zeta}_e {\vec v_e})
\over 1+\sqrt{1-v^2_e}} \Bigg]\Bigg\}. 
\label{A}
\end{equation}
One can see that the neutrino oscillation probability, $ P_{\nu_e
\rightarrow \nu_\mu}(x)$, the  mixing angle, $\theta_{eff}$, and
the oscillation length, $ L_{eff} $, exhibit dependence on the
total speed of electrons $\vec v_e$, correlation between $\vec
\beta$, $\vec v_e$ and polarization of matter $\vec \zeta_e$. The
resonance condition
\begin{equation}
 \Big(\delta m^{2}_\nu/2 |\vec p\,|\Big)\cos 2\theta= A,
\label{res}
\end{equation}
at which the probability has unit amplitude no matter how small
the vacuum mixing angle $\theta$ is, also depends on the motion and
polarization of matter and neutrino speed. It also follows that the
relativistic motion of matter could provide appearance of  the
resonance in the neutrino oscillations in certain cases when for
the given neutrino characteristics, $\delta m^{2}_\nu$, $|\vec
p\,|$ and $\theta$, and the invariant matter density at rest,
$n_0$, the resonance is impossible. Analysis of the neutrino effective 
potential in moving and polarized matter for different particular 
cases ( for different compositions, 
speeds and polarizations  of matter) can be found in 
ref. \cite {GriLobStuplb_02}.

We have developed the Lorentz invariant formalism for neutrino 
spin, flavour and spin-flavour oscillations in background matter 
and electromagnetic fields. Within this approach it becomes 
possible to describe neutrino spin oscillations in arbitrary 
electromagnetic fields. In particular, we predict new types of 
neutrino spin oscillations (and resonances) in the field of 
electromagnetic wave (e.m.w.) and the e.m.w. with the longitudinal 
magnetic field superimposed. We also generalized the Lorentz 
invariant approach to neutrino spin and flavour oscillations for 
the case of relativistic motion of background matter with effects 
of matter polarization are taken into account. It is shown that
the matter
terms in neutrino effective potentials depend on the values and
correlations of the three vectors, the neutrino and matter speeds
and matter polarization. In the case of relativistic motion of matter 
along (against) to 
neutrino propagation, matter effects in neutrino oscillations are
suppressed (increased). These effects can lead 
to interesting consequences for environments with  neutrino 
propagating through relativistic jets of matter.

I should like to thank Jean Tran Thanh Van for invitation 
to participate to the XXXVIIth Recontres de Moriond Session on Electroweak 
Interactions and Unified Theories and for support of my stay in Les Arcs.
I am also thankful to all the organizers for their hospitality.

\end{document}